\documentclass[showpacs,10pt,twocolumn,prb]{revtex4-1}
\usepackage{amsmath}
\usepackage{amssymb}
\usepackage{graphics}
\usepackage{epsfig}

\begin{document}

\title{Upper critical fields and superconducting anisotropy of K$_{0.70}$Fe$%
_{1.55}$Se$_{1.01}$S$_{0.99}$ and K$_{0.76}$Fe$_{1.61}$Se$_{0.96}$S$_{1.04}$
single crystals}
\author{Hechang Lei and C. Petrovic}
\affiliation{Condensed Matter Physics and Materials Science Department, Brookhaven
National Laboratory, Upton, NY 11973, USA}
\date{\today}


\begin{abstract}
We have investigated temperature and angular dependence of resistivity of K$_{0.70(7)}$Fe$_{1.55(7)}$Se$_{1.01(2)}$S$_{0.99(2)}$ and K$_{0.76(5)}$Fe$_{1.61(5)}$Se$_{0.96(4)}$S$_{1.04(5)}$ single crystals. The upper critical
fields $\mu _{0}H_{c2}(T)$ for both field directions decrease with the
increase in S content. On the other hand, the angle-dependent magnetoresistivity for both compounds can be
scaled onto one curve using the anisotropic Ginzburg-Landau theory. The
obtained anisotropy of $\mu _{0}H_{c2}(T)$ increases with S content,
implying that S doping might decrease the dimensionality of certain Fermi
surface parts, leading to stronger two dimensional character.
\end{abstract}

\maketitle

\section{Introduction}

Iron-based superconductors have stimulated intense activity after the
discovery of LaFeAsO$_{1-x}$F$_{x}$ with $T_{c}$ = 26 K (FeAs-1111 type)\cite%
{Kamihara}. Since then, the family of iron-based superconductors has been
gradually expanded to include AFe$_{2}$As$_{2}$ (A = alkaline or
alkaline-earth metals, FeAs-122 type)\cite{Rotter}, LiFeAs (FeAs-111 type)%
\cite{WangXC}, and $\alpha $-PbO type FeCh (Ch = S, Se, or Te, FeSe-11 type)%
\cite{HsuFC}. Even though these compounds share similar structural features
and possible same pairing s$\pm $ symmetry\cite{Mazin}, the superconducting
properties exhibit pronounced difference. Among these, the diversity of
temperature dependence of the upper critical field $\mu _{0}H_{c2}(T)$
attracts much interest since it provides valuable information on the
fundamental superconducting properties. These include coherence length,
anisotropy, details of underlying electronic structure, dimensionality of
superconductivity as well as insights into the pair-breaking mechanism. For
FeAs-1111 type and FeAs-122 type superconductors, the $\mu _{0}H_{c2}(T)$
can be described using a two-band model\cite{Hunte}$^{-}$\cite{Baily}. However,
for FeSe-11 type and arsenic-deficient FeAs-1111 type compounds, the $\mu
_{0}H_{c2}(T)$ exhibits Pauli-limiting behavior and satisfies the
single-band Werthamer-Helfand-Hohenberg (WHH) theory with strong
spin-paramagnetic effect and spin-orbital interaction\cite{LeiHC}$^{-}$\cite{Kida}.

Very recently, new family of iron-based superconductors A$_{x}$Fe$_{2-y}$Se$%
_{2}$ (A = K, Rb, Cs, and Tl, AFeSe-122 type) with maximum $T_{c}$\ $\approx
$ 33 K has been reported\cite{Guo}-\cite{FangMH}. Superconductivity in
AFeSe-122 materials is in proximity to an antiferromagnetic (AFM)
semiconducting state\cite{FangMH,WangDM}. This is different from other iron
based superconductors, which are usually close to spin density wave (SDW)
instability, and similar to cuprates where parent compounds are AFM Mott
insulators. Therefore, it is of interest to study whether this kind of
difference will lead to different behavior of $\mu _{0}H_{c2}(T)$.
Preliminary $\mu _{0}H_{c2,c}(T)$ and $\mu _{0}H_{c2,ab}(T)$ of AFeSe-122
estimated within simplified WHH model are about 40-70 T and 120-220 T,
respectively, giving the anisotropy of $\mu _{0}H_{c2}(0)$ about 3-4\cite%
{Mizuguchi4,YingJJ,LiCH,WangDM}. However, the $\mu _{0}H_{c2}(T)$ of K$_{x}$%
Fe$_{2-y}$Se$_{2}$ measured up to 60 T deviates from the WHH model and is
similar to FeAs-122 materials\cite{Mun}.

Insulating K$_{x}$Fe$_{2-y}$S$_{2}$ is isostructural with K$_{x}$Fe$_{2-y}$Se%
$_{2}$\cite{LeiHC3}. It is of interest to study changes of $\mu _{0}H_{c2}(T)
$ in AFeSe-122 with S doping. In this work, we report the upper critical
field anisotropy of K$_{0.70(7)}$Fe$_{1.55(7)}$Se$_{1.01(2)}$S$_{0.99(2)}$
and K$_{0.76(5)}$Fe$_{1.61(5)}$Se$_{0.96(4)}$S$_{1.04(5)}$ single crystals.
We show that both $\mu _{0}H_{c2,c}(T)$ and $\mu _{0}H_{c2,ab}(T)$ decrease
with S doping but the anisotropy of $\mu _{0}H_{c2}(T)$ is larger than in K$%
_{x}$Fe$_{2-y}$Se$_{2}$.

\section{Experiment}

The details of crystal growth and structure characterization are reported
elsewhere in detail\cite{LeiHC3,LeiHC5}. The average stoichiometry
determined by energy dispersive X-ray spectroscopy (EDX) was K:Fe:Se:S =
0.70(7):1.55(7):1.01(2):0.99(2) and 0.76(5):1.61(5):0.96(4):1.04(5) for S-99
and S-104, respectively. The in-plane resistivity $\rho _{ab}(T)$ was
measured using a four-probe configuration on rectangularly shaped and
polished single crystals with current flowing in the ab-plane of tetragonal
structure. Thin Pt wires were attached to electrical contacts made of Epotek
H20E silver epoxy. Sample dimensions were measured with an optical
microscope Nikon SMZ-800 with 10 $\mu $m resolution. Electrical transport
and magnetization measurements were carried out in a Quantum Design PPMS-9
and MPMS-XL5.

\section{Results and Discussion}

\begin{figure}[tbp]
\centerline{\includegraphics[scale=0.45]{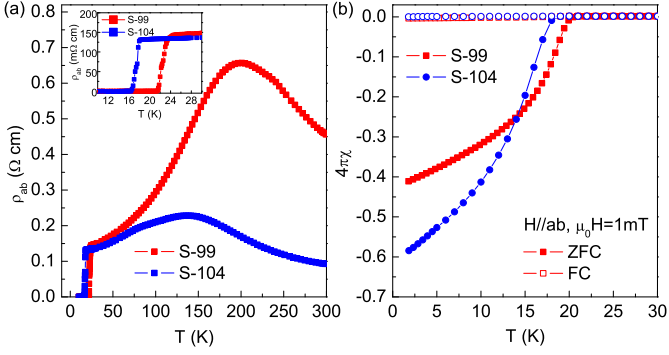}} \vspace*{-0.3cm}
\caption{(a) Temperature dependence of the in-plane resistivity $%
\protect\rho _{ab}(T)$ of S-99 and S-104 at zero field. Inset shows
resistivity near superconducting transition temperature. (b)
Temperature dependence of dc magnetic susceptibility of S-99 and S-104 with
ZFC and FC.}
\end{figure}

Fig.\ 1(a) shows the temperature dependence of in-plane resistivity $\rho
_{ab}(T)$ at zero field from 10 K to 300 K. Similar to K$_{x}$Fe$_{2-y}$Se$%
_{2}$, both S-99 and S-104 exhibit semiconducting behavior at high
temperature and then cross over to metallic behavior. With further decrease
in temperature, there are sharp superconducting transitions with $%
T_{c,onset} $ = 23.5(1) K and 18.0(2) K for S-99 and S-104, respectively. It
should be noted that the large shift of $T_{c}$ in our crystals can not be
explained by the small variations of K and Fe contents\cite{FangMH,WangDM}.
Fig. 1(b) shows the temperature dependence of the dc magnetic susceptibility
of S-99 and S-104 for $\mu _{0}H$ = 1 mT along the ab-plane. The
zero-field-cooling (ZFC) susceptibilities show that the superconducting
shielding emerges at about 18.0 K and 20.4 K for S-99 and S-104, consistent
with the $T_{c}$ obtained from resistivity measurement. The superconducting
volume fractions estimated from the ZFC magnetization at 1.8 K are about
0.41 and 0.58 for S-99 and S-104, respectively, indicating substantial,
albeit still filamentary superconducting volume fraction.

\begin{figure}[tbp]
\centerline{\includegraphics[scale=0.45]{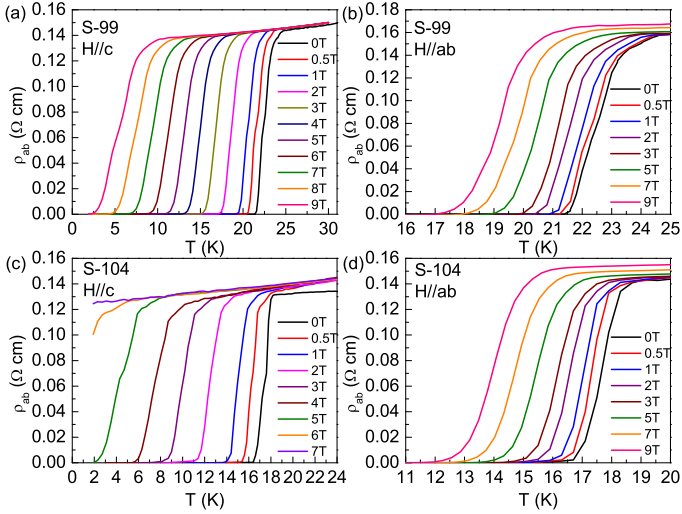}} \vspace*{-0.3cm}
\caption{Temperature dependence of the resistivity $\protect\rho _{ab}(T)$
in magnetic fields for (a) H$\Vert $ab and (b) H$\Vert $c
of S-99 and (c) H$\Vert $ab and (d) H$\Vert $c of \ S-104.}
\end{figure}

Fig. 2 shows temperature dependent resistivity of $\rho _{ab}(T)$ of S-99
and S-104 in magnetic fields up to 9 T for H$\Vert $c and H$\Vert $ab. With
increasing magnetic fields, the superconducting transitions shift to lower
temperature gradually and transition widths become broader. This trend is
more obvious for H$\Vert $c than for H$\Vert $ab. This is similar to
Fe(Te,Se) and Fe(Te,S) single crystals\cite{LeiHC,LeiHC2}. When compared to
pronounced broadening of resistivity in magnetic field in FeAs-1111
compounds for H$\Vert $c\cite{LeeHS,Karpinski}, the broadening of
resistivity in AFeSe-122 materials is far smaller for both field directions,
similar to the FeAs-122 and FeSe-11 compounds\cite%
{LeiHC,LeiHC2,WangZS,Bukowski}. This indicates that the vortex-liquid state
region should be narrower in AFeSe-122. It should be noted that this narrow transition widths could also have contribution from the normal state parts of samples. On the other hand, the superconductivity in
S-99 and S-104 is more sensitive to the field for H$\Vert $c when compared
to undoped K$_{x}$Fe$_{2-y}$Se$_{2}$. The $T_{c,onset}$ of S-99 has shifted
to 7.6(2) K at $\mu _{0}H$ = 9 T, whereas for S-104, the superconductivity
has been\ suppressed completely above 1.9 K even at $\mu _{0}H$ = 7 T. This
indicates that the $\mu _{0}H_{c2,c}(T)$ of S-99 and S-104 are much lower
than in K$_{x}$Fe$_{2-y}$Se$_{2}$.

\begin{figure}[tbp]
\centerline{\includegraphics[scale=0.45]{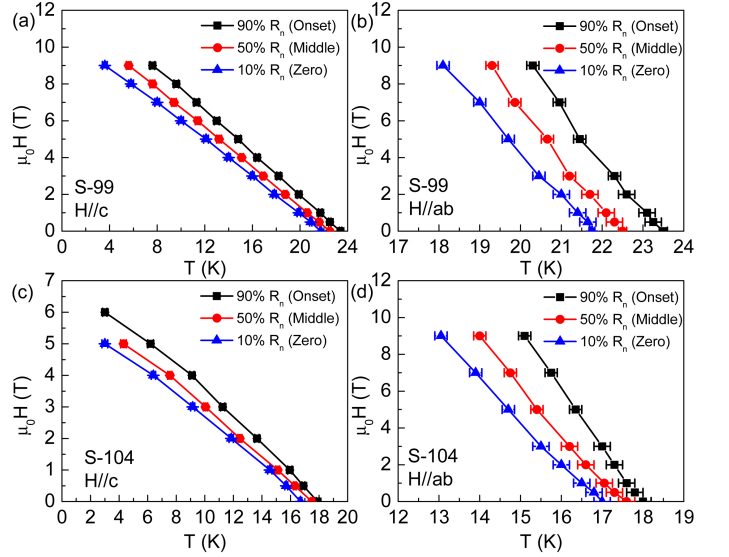}} \vspace*{-0.3cm}
\caption{Temperature dependence of the resistive upper critical field $%
\protect\mu _{0}H_{c2}(T)$ of S-99 for (a) H$\Vert $ab and (b) H$\Vert $c, and of S-104 for (c) H$\Vert $ab and (d) H$\Vert $%
c.}
\end{figure}

Fig. 3 presents the temperature dependence of the resistive upper critical
fields $\mu _{0}H_{c2}(T)$ of S-99 and S-104 determined from resistivity
drops to 90\% (Onset), 50\% (Middle) and 10\% (Zero) of the normal state
resistivity $\rho _{n,ab}(T,H)$ for both field directions. The normal-state
resistivity $\rho _{n,ab}(T,H)$ was determined by linearly extrapolating the
normal-state behavior above the onset of superconductivity in $\rho
_{ab}(T,H)$ curves. It can be seen that the $\mu _{0}H_{c2}(T)$ for H$\Vert $%
c is much smaller than that for H$\Vert $ab. Hence the low field anisotropy of $\mu
_{0}H_{c2}(T)$ for both of S-99 and S-104 is large, similar to K$_{x}$Fe$%
_{2-y}$Se$_{2}$\cite{Mizuguchi4,YingJJ,WangDM}. On the other hand, the low field positive curvature of $\mu
_{0}H_{c2,ab}(T)$ shows a slight positive curvature which may be due to a crossover from three dimensions (3D) to two dimensions (2D)\cite{Prober}.

According to the conventional single-band WHH theory which describes the
orbital limited upper critical field of dirty type-II superconductors\cite%
{Werthamer}, the $\mu _{0}H_{c2}$ can be described by

\begin{equation}
\ln \frac{1}{t}\text{=}\psi (\frac{1}{2}+\frac{\overset{-}{h}}{2t})-\psi (%
\frac{1}{2})
\end{equation}%
where $t=T/T_{c}$, $\psi $ is a digamma function and

\begin{equation}
\overset{-}{h}\text{=}\frac{4H_{c2}}{\pi ^{2}T_{c}(-dH_{c2}/dT)_{T=T_{c}}}
\end{equation}

\begin{figure}[tbp]
\centerline{\includegraphics[scale=0.45]{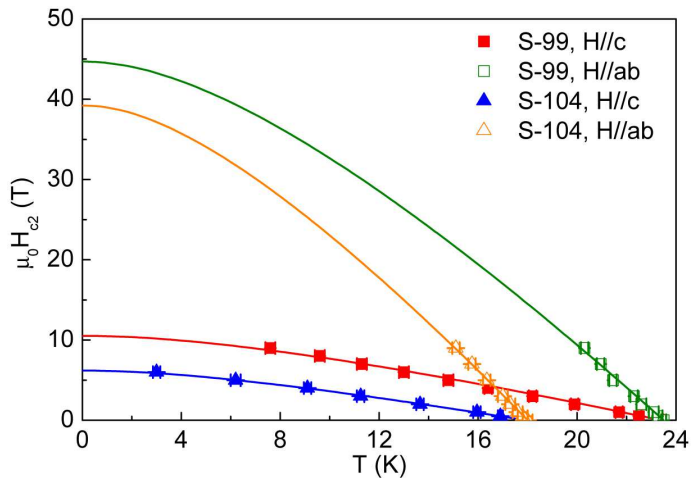}} \vspace*{-0.3cm}
\caption{Fits to $\protect\mu _{0}H_{c2,c}(T)$ and $\protect\mu _{0}H_{c2,ab}(T)$ using the simplified WHH
theory for (a) S-99 and (b) S-104. The $\protect\mu _{0}H_{c2,c}(T)$ and $\protect\mu _{0}H_{c2,ab}(T)$ of S-99
and S-104 are determined from 90\% of $\protect\rho_{n,b}(T,H)$.}
\end{figure}

\begin{table*}[tbp]
\caption{Superconducting parameters of S-99 and S-104 single crystals.}
\label{TableKey}\centering%
\begin{tabular}{cccccccc}
\hline\hline
&  & $T_{c,onset}(\mu _{0}H=0)$ & $-d(\mu _{0}H_{c2})/dT|_{T=T_{c}}$ & $%
H_{c2}(0)$ & $\Gamma (0)(=H_{c2,ab}/H_{c2,c})$ & $\xi _{ab}(0)$ & $\xi
_{c}(0)$ \\
&  & (K) & (T/K) & (T) &  & (nm) & (nm) \\ \hline
S-99 & H$\parallel $ab & 23.5(1) & 2.74(7) & 44.69 & 4.25 & 5.60 & 1.32 \\
& H$\parallel $c & 23.4(1) & 0.649(7) & 10.51 &  &  &  \\ \hline
S-104 & H$\parallel $ab & 18.0(2) & 3.15(4) & 39.22 & 6.34 & 7.29 & 1.15 \\
& H$\parallel $c & 17.9(1) & 0.499(4) & 6.19 &  &  &  \\ \hline\hline
\end{tabular}%
\end{table*}

Using the measured $T_{c,onset}$ and setting the slopes of $\mu
_{0}H_{c2}(T) $ near $T_{c,onset}(\mu _{0}H=0)$, $-d(\mu
_{0}H_{c2})/dT|_{T=T_{c}}$, as free parameters, the $\mu _{0}H_{c2}(0)$ of
S-99 and S-104 along both field directions can be obtained using the
simplified WHH model. As shown in Table 1, the fitted $-d(\mu
_{0}H_{c2})/dT|_{T=T_{c}}$ are much smaller than in K$_{x}$Fe$_{2-y}$Se$_{2}$%
, which are about 2-3 T/K for H$\Vert $c and 6-10 T/K for H$\Vert $ab\cite%
{WangDM,YingJJ}. Correspondingly, the $\mu _{0}H_{c2}(0)$ for both field
directions are also much smaller than in K$_{x}$Fe$_{2-y}$Se$_{2}$\cite%
{Mizuguchi4,YingJJ,WangDM,Mun}. From the $\mu _{0}H_{c2}(0)$,
zero-temperature coherence length $\xi (0)$ can be estimated with
Ginzburg-Landau (GL) formula $\mu _{0}H_{c2,c}(0)$=$\Phi _{0}/[2\pi \xi
_{ab}^{2}(0)]$, and $\mu _{0}H_{c2,ab}(0)$=$\Phi _{0}/[2\pi \xi _{ab}(0)\xi
_{c}(0)]$ where $\Phi _{0}$=2.07$\times $10$^{-15} $ $Wb$. The
zero-temperature anisotropy $(\Gamma (0)$ $=H_{c2,ab}(0)/H_{c2,c}(0))$
obtained from WHH fits for S-99 and S-104 is 4.26 and 5.56, respectively.
The $\Gamma (0)$ of S-104 is larger than that of S-99, and both are larger
than in K$_{x}$Fe$_{2-y}$Se$_{2}$\cite{Mizuguchi4,YingJJ,WangDM}. All
obtained parameters are listed in Table 1 and the fitting results are show in fig. 4.

It should be noted that the $\mu _{0}H_{c2,c}(T)$ of S-99 can be fitted linearly with slightly lower fitting error when compared to simplified WHH theory, whereas the fitting quality using simplified WHH theory is much better than linear function for S-104. This linear behavior has also been observed in K$_{x}$Fe$_{2-y}$Se$_{2}$\cite{Mun}. It implies that S doping could change the temperature dependence of $\mu _{0}H_{c2,c}(T)$ possibly due to the changes of band structure. Agreement with simplified WHH theory with temperature far below $%
T_{c}$ for H$\Vert $c implies that for S-99 the orbital effect
should be the dominant pair-breaking mechanism and spin-paramagnetic effect and spin-orbital interaction could be negligible when the magnetic field is
applied along c axis. This is different from FeSe-11
materials where the\ $\mu _{0}H_{c2,c}(T)$ exhibits Pauli-limiting behavior
and strong spin-orbital interaction has to be considered\cite{LeiHC,LeiHC2}. Pauli limiting field is $\mu _{0}H_{p}(0)$ = 1.86$T_{c}(1+\lambda
_{e-ph})^{1/2}$, where $\lambda _{e-ph}$ is electron-phonon coupling
parameter\cite{Orlando}. Using the typical value for weak-coupling BCS
superconductors ($\lambda _{e-ph}$ = 0.5)\cite{Allen}, we obtain $\mu
_{0}H_{p}(0)$ = 53.5 T and 41.0 T for S-99 and S-104, respectively. Both
values are larger than extrapolated $\mu _{0}H_{c2,c}(0)$ using simplified WHH theory for S-99 and S-104 and also larger than the value determined from linear extrapolation ($\mu _{0}H_{c2,c}(0)$ = 13.47(4) T) for S-99. On the other hand, the $\mu _{0}H_{c2,c}(T)$ in measured region cannot be fitted using two-band model, which is different from FePn-1111 and FeAs-122 materials\cite{Hunte}-\cite{Baily}. Experiments in higher field and lower temperature are needed in order to shed more light on the upper critical field behavior for both magnetic field directions.

\begin{figure}[tbp]
\centerline{\includegraphics[scale=0.85]{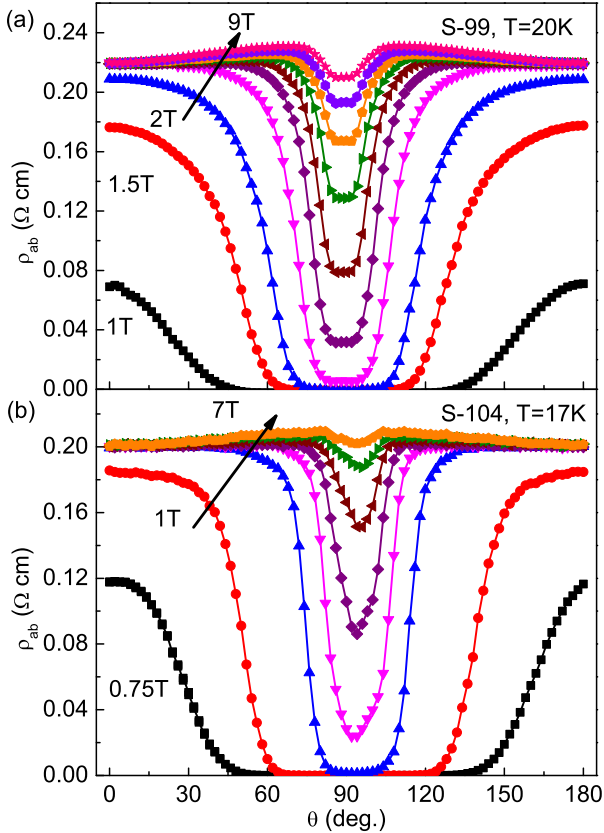}} \vspace*{-0.3cm}
\caption{Angular dependence of $\protect\rho _{ab}(%
\protect\theta, \mu_{0}H)$ for (a) S-99 at 20 K with $\mu_{0}H$ = 1, 1.5, 2, 3, 4, 5, 6, 7, 8, and 9 T and (b) S-104 at
17 K with $\mu_{0}H$ = 0.75, 1, 2, 3, 4, 5, 6, and 7 T.}
\end{figure}

There is uncertainty in estimated anisotropy ratio, connected with the
uncertainty in determining the upper critical field values from resistivity.
This can be solved to some extent by the measurements of angular-dependent
resistivity $\rho _{ab}(\theta, \mu_{0}H)$. Fig. 5 (a) and (b) shows the
angular-dependent resistivity for S-99 at 20 K and S-104 at 17 K,
respectively. All curves have common cup-like shape and the minimum value at
$\theta =90^{\circ }$, where $\theta $ is the angle between the direction of
external filed and the c axis. This indicates that the upper critical field
along the ab plane is larger than along the c axis for both samples.
According to the anisotropic GL model, the effective upper critical field $%
H_{c2}^{GL}$($\theta $) can be represented as\cite{Blatter}

\begin{equation}
\mu _{0}H_{c2}^{GL}(\theta )=\mu _{0}H_{c2,ab}/(\sin ^{2}\theta +\Gamma
^{2}\cos ^{2}\theta )^{1/2}
\end{equation}

where $\Gamma =H_{c2,ab}/H_{c2,c}=(m_{c}/m_{ab})^{1/2}=\xi _{ab}/\xi _{c}$.

Since the resistivity in the mixed state depends on the effective field $%
H/H_{c2}^{GL}$($\theta $), the resistivity can be scaled with $H/H_{c2}^{GL}$%
($\theta $) and should collapse onto one curve in different magnetic fields
at a certain temperature when a proper $\Gamma (T)$ value is chosen\cite{Blatter2}. Fig. 6 shows the relation between resistivity and scaling field $\mu _{0}H_{s}=\mu _{0}H(\sin ^{2}\theta +\Gamma ^{2}\cos ^{2}\theta )^{1/2}$%
. It can be seen that by adjusting $\Gamma (T)$, a good scaling behavior for
S-99 and S-104 can be obtained. The determined $\Gamma (T)$ are shown in the insets of fig. 6(a) and (b). The values are similar to those obtained from fig. 3. But because only one fitting parameter $\Gamma
(T)$ can be adjusted for scaling at each temperature, the obtained value of $%
\Gamma (T)$ is more reliable than that determined from the ratio of $\mu
_{0}H_{c2,ab}(T)$ to $\mu _{0}H_{c2,c}(T)$ (which may be influenced by
difference among onset, middle and zero resistivity as well as possible misalignment of field). For both crystals, the
$\Gamma (T)$ determined for GL theory exhibits the same trend. Anisotropy increases with decreasing
temperature from 22 K to 21 K for S-99 and from 17 K to 16 K for S-104 and
then almost unchanged (as shown in the insets of Fig. 6 (a) and (b)). The
similar behavior is also observed in pure K$_{x}$Fe$_{2-y}$Se$_{2}$\cite{Mun}. It should be noted that the anisotropy increases with S doping. It changes from $\sim $ 3 for pure K$_{x}$Fe$_{2-y}$Se$_{2}$ to $\sim $ 6 for
S-104\cite{Mizuguchi4,YingJJ,WangDM,Mun}. The larger anisotropy with increasing S content may suggest that two
dimensional Fermi surface is becoming less warped with S doping\cite{LiCH}.

\begin{figure}[tbp]
\centerline{\includegraphics[scale=0.45]{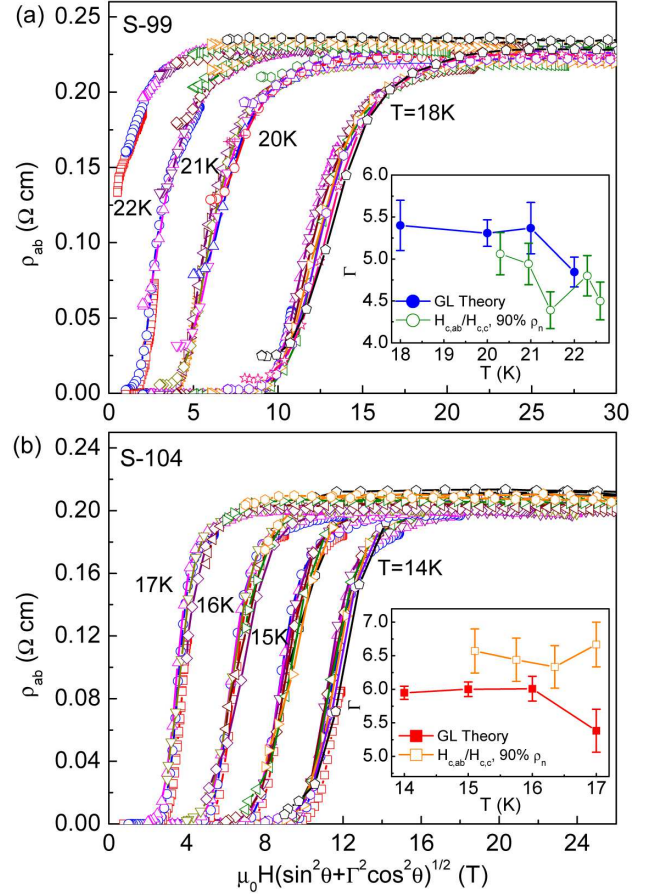}} \vspace*{-0.3cm}
\caption{Scaling behavior of the resistivity versus $\protect\mu _{0}H_{s}=%
\protect\mu _{0}H(\cos ^{2}\protect\theta +\Gamma ^{2}\sin ^{2}\protect%
\theta )^{1/2}$ for (a) S-99 and (b) S-104 at different
magnetic fields and temperatures. The inset (a) and (b)
show the temperature dependence of $\Gamma (T)$ determined using GL theory and from fig. 3 for S-99 and
S-104, respectively.}
\end{figure}

\section{Conclusion}

In conclusion, we have investigated the upper critical fields of K$%
_{0.70(7)} $Fe$_{1.55(7)}$Se$_{1.01(2)}$S$_{0.99(2)}$ and K$_{0.76(5)}$Fe$%
_{1.61(5)}$Se$_{0.96(4)}$S$_{1.04(5)}$ single crystals. When compared to
pure K$_{x}$Fe$_{2-y}$Se$_{2},$ it is found that the $\mu _{0}H_{c2}(T)$
decreases with the increase S content for both field directions. Moreover, the temperature dependence of $\mu _{0}H_{c2,c}(T)$ indicates that
spin-paramagnetic effect and spin-orbital interaction could be negligible for H$\parallel $c. On the other hand, it
is found that angular-dependence of resistivity $\rho _{ab}(\theta ,H)$
follows a scaling law based on the anisotropic GL theory. The values of mass
tensor anisotropy $\Gamma (T)$ increase with increasing S content.

\section{Acknowledgements}

We thank John Warren for help with scanning electron microscopy
measurements. Work at Brookhaven is supported by the U.S. DOE under Contract
No. DE-AC02-98CH10886 and in part by the Center for Emergent
Superconductivity, an Energy Frontier Research Center funded by the U.S.
DOE, Office for Basic Energy Science.

\end{document}